\documentclass[12pt,rmp,twocolumn,natbib,showkeys,pre,floatfix]{revtex4}

\usepackage[dvips]{graphicx}

\begin{document}
\title{Verhulst model with L\'{e}vy white noise excitation}

\author{\firstname{A. A.} \surname{Dubkov$^{\sharp}$}}
\author{\firstname{B.} \surname{Spagnolo$^{\star}$}}
\affiliation{$^{\sharp}$ Radiophysics Faculty, Nizhniy Novgorod
State University \ \\23 Gagarin Ave., 603950 Nizhniy Novgorod,
Russia\footnote{e-mail: dubkov@rf.unn.ru}} \affiliation{$^{\star}$
Dipartimento di Fisica e Tecnologie Relative and CNISM-INFM,
\\ Group of Interdisciplinary
Physics\footnote{http://gip.dft.unipa.it}, Universit$\grave{a}$ di
Palermo, \\ Viale delle Scienze, I-90128, Palermo,
Italy\footnote{e-mail: spagnolo@unipa.it}}

\begin{abstract}
The transient dynamics of the Verhulst model perturbed by arbitrary
non-Gaussian white noise is investigated. Based on the infinitely
divisible distribution of the L\'{e}vy process we study the
nonlinear relaxation of the population density for three cases of
white non-Gaussian noise: (i) shot noise, (ii) noise with a
probability density of increments expressed in terms of Gamma
function, and (iii) Cauchy stable noise. We obtain exact results for
the probability distribution of the population density in all cases,
and for Cauchy stable noise the exact expression of the nonlinear
relaxation time is derived. Moreover starting from an initial delta
function distribution, we find a transition induced by the
multiplicative L\'{e}vy noise, from a trimodal probability
distribution to a bimodal probability distribution in asymptotics.
Finally we find a nonmonotonic behavior of the nonlinear relaxation
time as a function of the Cauchy stable noise intensity.

\end{abstract}

\date{\today}

\maketitle

\section{Introduction}

The nonlinear stochastic systems with noise excitation have
attracted extensive attention and the concept of noise-induced
transitions has got a wide variety of applications in physics,
chemistry, and biology~\cite{Hor84}. Noise-induced transitions are
conventionally defined in terms of changes in the number of extrema
in the probability distribution of a system variable and may depend
both quantitatively and qualitatively on the character of the noise,
i.e. on the properties of stochastic process which describes the
noise excitation. The Verhulst model, which is a cornerstone of
empirical and theoretical ecology, is one of the classic examples of
self-organization in many natural and artificial
systems~\cite{Eig79}. This model, also known as the logistic model,
is relevant to a wide range of situations including population
dynamics~\cite{Hor84,Mor82,Ciu93,Mat00}, self-replication of
macromolecules~\cite{Eig71}, spread of viral epidemics~\cite{Ace06},
cancer cell population \cite{Bao03}, biological and biochemical
systems~\cite{Der90,Ciu96}, population of photons in a single mode
laser~\cite{McN74,Oga83}, autocatalytic chemical
reactions~\cite{Sch72,Cha76,Gar77,Bou82,Leu87}, freezing of
supercooled liquids~\cite{Das83}, social
sciences~\cite{Her72,Mon78}, etc.

In considering how the population density $x\left(t\right)$ may
change with time $t$, Verhulst proposed the following equation
\begin{equation}
\frac{dx}{dt}=rx\left( 1-\frac{x}{\Omega }\right) .
\label{F-1}
\end{equation}
where there is the Malthus term with the rate constant $r$ and a
saturation term with the $\Omega$ factor, which is the upper limit
for the population growth due to the availability of the resources.

Really the parameters $r$ and $\Omega$ are not constant. In fact the
parameter $r$ changes randomly due to season fluctuations, and the
parameter $\Omega$ fluctuates due to the environmental interaction
which causes the random availability of resources. As a consequence
we have the following stochastic Verhulst equation
\begin{equation}
\frac{dx}{dt}=r\left( t\right) x\left[ 1-\frac{x}{\Omega
\left(t\right) }\right] .
\label{F-2}
\end{equation}
In the context of macromolecular self-replication, the model
equation~(\ref{F-2}), with constant $\Omega$ and a white Gaussian
noise in $r\left( t\right)$, was numerically studied in
Ref.~\cite{Leu88} and the critical slowing down, i.e. a divergence
of the relaxation time at some noise intensity, was found. Later
Jackson and co-authors~\cite{Jac89} investigated the same model, by
analog experiment and digital simulations. They analyzed
specifically in detail the nonlinear relaxation time defined
as~\cite{Bin73}
\begin{equation}
T=\frac{\int\nolimits_{0}^{\infty}\left[\left\langle x\left(
t\right) \right\rangle -\left\langle x\left(\infty\right)
\right\rangle \right] dt}{x\left(0\right) - \left\langle x\left(
\infty\right) \right\rangle}
\label{F-3}
\end{equation}
and did not observe the critical slowing down. They explained this
discrepancy by the incorrect approximate truncation of the
asymptotic power series for $T$ used in Ref.~\cite{Leu88}. The
stability conditions were derived in Ref.~\cite{Gol03}. Similar
investigations for colored Gaussian noise $r(t)$ were performed in
Ref.~\cite{Man90}, where a monotonic dependence of the relaxation
time and the correlation time on the noise intensity was found. The
generalization of Eq.~(\ref{F-2}), to study a
Bernoulli-Malthus-Verhulst model driven by a multiplicative colored
noise, was analyzed recently in
Ref.~\cite{Cal07}.\\
\indent The evolution of the mean value in the case of
Eq.~(\ref{F-2}) with constant $r$ and white Gaussian noise
excitation $\beta\left( t\right) =r/\Omega \left( t\right)$ was
considered in Refs.~\cite{Ciu93,Suz82,Suz82a,Bre82,Mak85,Mor86}. In
Refs.~\cite{Mak85,Mor86} the authors, using perturbation technique,
obtained the exact expansion in power series on noise intensity of
all the moments and found the long-time decay of $t^{-1/2}$. In
Ref.~\cite{Ciu93} the authors derived the long-time behavior of all
the moments of the population density by means of an exact
asymptotic expansion of the time averaged process generating
function, and found the same asymptotic behavior of $t^{-1/2}$ at
the critical point. This very slow relaxation of the moments near
the critical point is the phenomenon of critical slowing down.\\
\indent In the present paper, using the previously obtained results
for a generalized Langevin equation with a L\'{e}vy noise
source~\cite{Dub05,Dub08}, we investigate the transient dynamics of
the stochastic Verhulst model with a fluctuating growth rate and a
constant value for the saturation population density $\Omega$, that
is $\Omega=1$. The exact results for the mean value of the
population density and its nonstationary probability distribution
for different types of white non-Gaussian excitation $r\left(
t\right)$ are obtained. We find the interesting noise-induced
transitions for the probability distribution of the population
density and the relaxation dynamics of its mean value for Cauchy
stable noise. Finally we obtain a nonmonotonic behavior of the
nonlinear relaxation time as a function of the Cauchy noise
intensity.

\section{Stochastic Verhulst equation with non-Gaussian fluctuations of growth
rate}

Let us consider Eq.~(\ref{F-2}) with a constant saturation value
$\Omega=1$, namely
\begin{equation}
\frac{dx}{dt}=r\left(  t\right)  x\left(  1-x\right)  .
\label{F-4}
\end{equation}
After changing variable $y=\ln[x/(1-x)]$, we obtain
\[
y\left(t\right) = y\left(0\right) +\int\nolimits_{0}^{t}r\left(
\tau\right) d\tau \,
\]
and the exact solution of Eq.~(\ref{F-4}) is
\begin{equation}
x\left(t\right) = \left(1+\frac{1-x_{0}}{x_{0}}\exp\left\{ -\int
\nolimits_{0}^{t}r\left(\tau\right) d\tau\right\} \right)^{-1},
\label{F-5}
\end{equation}
where $x_{0}=x\left(0\right)$. Now by substituting in
Eq.~(\ref{F-5}) the following expression for the random rate $r(t)$
\begin{equation}
r\left( t\right) =r+\xi\left( t\right),
\label{F-6}
\end{equation}
where $r>0$ and $\xi\left( t\right)$ is an arbitrary white
non-Gaussian noise with zero mean, we can rewrite the
solution~(\ref{F-5}) as
\begin{equation}
x\left(t\right) =
\left(1+\frac{1-x_{0}}{x_{0}}e^{-rt-L\left(t\right)}\right)^{-1}.
\label{F-7}
\end{equation}
Here $L\left(t\right)$ denotes the so-called L\'{e}vy random process
with $L\left(0\right)=0$, and
$\xi\left(t\right)=\dot{L}\left(t\right) $. As it was shown in
Refs.~\cite{Dub05,Dub08,Fel71}, L\'{e}vy processes having stationary
and statistically independent increments on non-overlapping time
intervals belongs to the class of stochastic processes with
infinitely divisible distributions. As a consequence, the
characteristic function of $L\left( t\right)$ can be represented in
the following form (see Eq.~(6) in ~\cite{Dub05})
\begin{equation}
\left\langle e^{iuL\left(  t\right)  }\right\rangle =\exp\left\{
t\int\nolimits_{-\infty}^{+\infty}\frac{e^{iuz}-1-iu\sin
z}{z^{2}}\rho\left( z\right)  dz\right\}  ,
\label{F-8}
\end{equation}
where $\rho\left(z\right)$ is some non-negative kernel function. The
case $\rho\left(z\right)=2D\delta\left(  z\right)$ corresponds to a
white Gaussian noise excitation $\xi\left(t\right)$, while for a
symmetric L\'{e}vy stable noise $\xi\left( t\right)$ with index
$\alpha$ we have a power-law kernel $\rho\left( z\right)=Q\left\vert
z\right\vert ^{1-\alpha}$, with $0<\alpha<2$.

\begin{figure}[ptbh]
\centering{\resizebox{7cm}{!}{\includegraphics{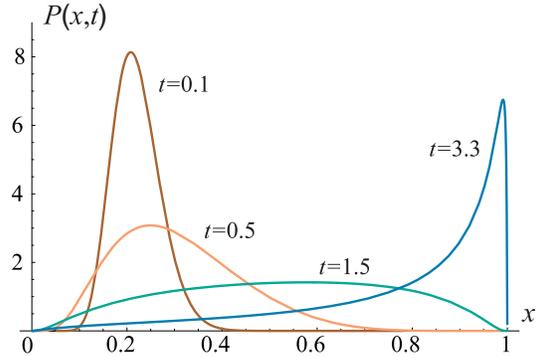}}}
\caption{Time evolution of the probability distribution of the
population density for white Gaussian noise excitation with
intensity $D$. The values of the parameters are: $x_0=0.2$, $r=1$,
$D=0.5$.} \label{fig-1}
\end{figure}

In the model under consideration the stationary probability
distribution has: (i) a singularity at the stable point $x = 1$ for
white Gaussian noise, and (ii) two singularities at both stable
points $x=0$ and $x=1$ for L\'{e}vy noise. To analyze the time
behavior of the probability distribution in the transient dynamics
it is better not to use the Kolmogorov equation for the probability
density $P\left(x,t\right)$, but rather the exact
solution~(\ref{F-7}). Using the standard theorem of the probability
theory regarding a nonlinear transformation of a random variable, we
find from Eq.~(\ref{F-7})
\begin{equation}
P\left(x,t\right) =\frac{1}{x\left(1-x\right)}P_{L}\left(\ln
\left[\frac{\left(1-x_{0}\right)x}{x_{0}\left(1-x\right)}\right]
-rt,t\right),
\label{F-9}
\end{equation}
where $P_{L}\left(z,t\right)$ is the probability density
corresponding to the characteristic function~(\ref{F-8}). For a
white Gaussian noise $\xi\left(t\right)$, this distribution reads
\begin{equation}
P_{L}\left(  z,t\right)  =\frac{1}{2\sqrt{\pi Dt}}\exp\left\{
-\frac{z^{2} }{4Dt}\right\}  .
\label{F-10}
\end{equation}
The time evolution of the probability distribution
$P\left(x,t\right) $ for $D=0.5$, $r=1$, and $x_{0}=0.2$ is plotted
in Fig.~\ref{fig-1}.

As it is easily seen, the maximum of the unimodal distribution with
initial position at $x=0.2$ shifts with time towards the stable
point at $x=1$. At the same time, as it follows from
Eqs.~(\ref{F-9}) and~(\ref{F-10}), for all $t>0$ we have
\begin{equation}
\lim_{x\rightarrow 0^+}P\left(x,t\right) = \lim_{x\rightarrow
1^-}P\left(x,t\right) = 0.
\label{F-11}
\end{equation}

\begin{figure}[ptbh]
\centering{\resizebox{7cm}{!}{\includegraphics{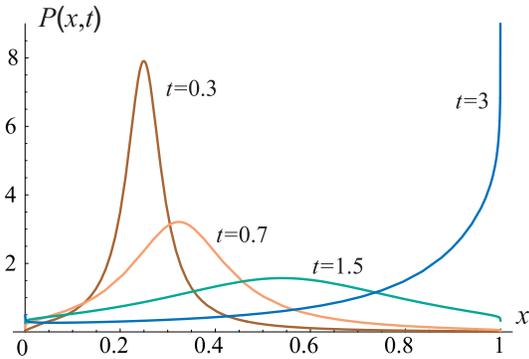}}}
\vskip-0.4cm\caption{Time evolution of the probability distribution
of the population density in the case of L\'{e}vy noise with
distribution~(\ref{F-12}). The values of the parameters are:
$x_0=0.2$, $r=1$, $K=0.43$.} \label{fig-2}
\end{figure}

The same picture is observed for another kernel function
$\rho\left(z\right) = Kz/\left(2\sinh z\right)$ $\left(K>0\right)$,
corresponding to a L\'{e}vy process $\eta\left(t\right)$ with finite
moments and the following probability density of increments
\begin{equation}
P_{L}\left(  z,t\right)
=\frac{2^{Kt-1}}{\pi^{2}\Gamma\left(Kt\right)} \Gamma\left(
\frac{Kt}{2}+\frac{iz}{\pi}\right) \Gamma\left(\frac{Kt}
{2}-\frac{iz}{\pi}\right)  ,
\label{F-12}
\end{equation}
where $\Gamma\left(x\right)$ is the Gamma function. The
corresponding time evolution of the probability distribution
$P\left( x,t\right)$ for $K=0.43$, $r=1$, and $x_{0}=0.2$ is shown
in Fig.~\ref{fig-2}.

\begin{figure}[ptbh]
\centering{\resizebox{7cm}{!}{\includegraphics{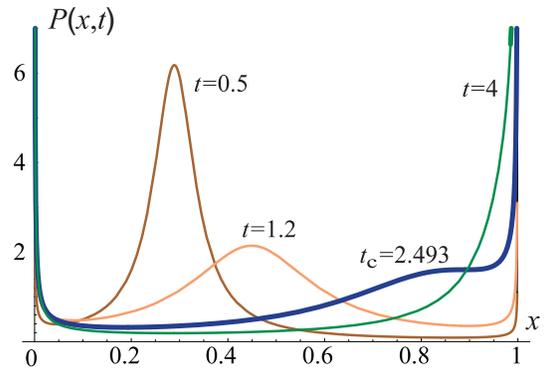}}}
\caption{Time evolution of the probability distribution of the
population density in the case of white Cauchy noise excitation. The
values of the parameters are: $x_0 = 0.2$, $r = 1$, $D_1 = 0.5$.}
\label{fig-3}
\end{figure}

\begin{figure}[ptbh]
\centering{\resizebox{7cm}{!}{\includegraphics{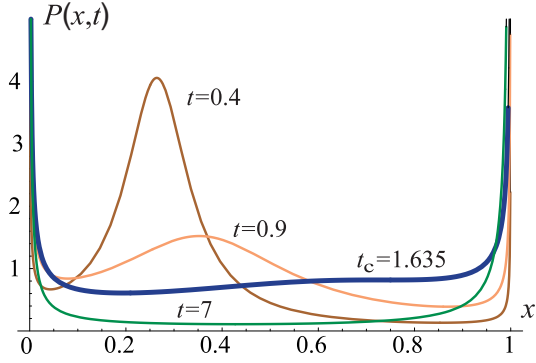}}}
\caption{Time evolution of the probability distribution of the
population density in the case of white Cauchy noise. The values of
the parameters are $x_0=0.2$, $r=1$, $D_1=1$.} \label{fig-4}
\end{figure}

\begin{figure}[ptbh]
\centering{\resizebox{7cm}{!}{\includegraphics{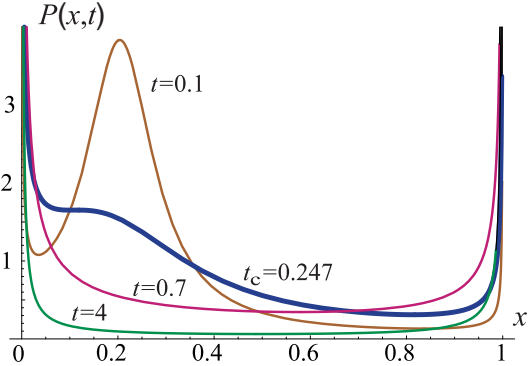}}}
\caption{Time evolution of the probability distribution of the
population density in the case of white Cauchy noise. The values of
the parameters are $x_0=0.2$, $r=1$, $D_1=5$.} \label{fig-5}
\end{figure}

A different situation we have for a Cauchy stable noise
$\xi\left(t\right)$ with constant kernel $\rho\left(z\right) = Q$
$\left( \alpha=1\right)$. After evaluation of the integral in
Eq.~(\ref{F-8}), the probability density of the L\'{e}vy process
increments takes the form of the well-known Cauchy
distribution~\cite{Fel71}
\begin{equation}
P_{L}\left(z,t\right) = \frac{D_{1}t}{\pi\left[z^{2}+
\left(D_{1}t\right)^{2}\right]},
\label{F-13}
\end{equation}
where $D_{1}=\pi Q$ is the noise intensity parameter. In such a case
from Eqs.~(\ref{F-9}) and~(\ref{F-13}) for all $t>0$ we find
\begin{equation}
\lim_{x\rightarrow 0^+}P\left(  x,t\right) =\lim_{x\rightarrow
1^-}P\left( x,t\right)  = \infty.
\label{F-14}
\end{equation}
As a result, from an initial delta function we immediately obtain a
trimodal distribution for $t>0$ and then after some transition time
$t_c$ a bimodal one with two singularities at the stable points
$x=0$ and $x=1$ (see Figs.~\ref{fig-3},~\ref{fig-4}
and~\ref{fig-5}). We should note that the transition from trimodal
to bimodal distribution is a general feature of the model in the
presence of a Cauchy stable noise, and it is not limited to some
range of parameters. In fact, from Eq.~(\ref{F-14}) and a delta
function initial distribution inside the interval (0,1), this
transition always takes place.
In the following Figs.~\ref{fig-4} and~\ref{fig-5} we show the time
evolution of the probability distribution of the population density
for two other values of the noise intensity, namely $D_1 = 1$ and
$D_1 = 5$. As the noise intensity increases the probability
distribution shows two singularities near $x = 0$ and $x = 1$ with
different amplitude.

This transition in the shape of the probability distribution of the
population density is due to both the multiplicative noise and the
L\'{e}vy noise source. Using Eqs.~(\ref{F-9}) and~(\ref{F-13}) and
equating to zero the derivative of $P(x,t)$ with respect to $x$, we
obtain the following condition for the extrema in the range $0<x<1$,
and particularly for a minimum in the same interval
\begin{equation}
\frac{z(x,t)}{z(x,t)^2 + (D_1 t)^2} = x - \frac{1}{2} \; ,
\label{F-15}
\end{equation}
with
\begin{equation}
z(x,t) = \ln
\left[\frac{\left(1-x_{0}\right)x}{x_{0}\left(1-x\right)}\right] -rt
\,.
\label{F-16}
\end{equation}
This condition can be solved graphically by finding the intersection
between the functions $y_1 = z(x,t)/(z(x,t)^2 + (D_1 t)^2)$ and $y_2
= x - 1/2$. This is done in the following
Figs.~\ref{fig-6},~\ref{fig-7},~\ref{fig-8}, where the function
$y_1$ is plotted for three different values of time and noise
intensity. In each figure the black blue curve (color on line)
corresponds to the critical value of time $t_c$ for which we have a
noise induced transition of the probability distribution of the
population density from trimodal to bimodal, that is from two minima
and one maximum to one minimum inside the interval $0 < x < 1$. The
appearance of one minimum in the probability distribution is the
signature of this transition.
\begin{figure}[ptbh]
\centering{\resizebox{7cm}{!}{\includegraphics{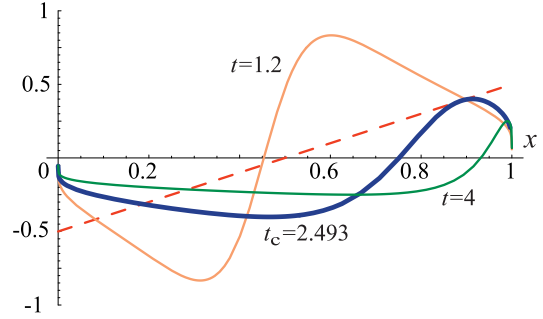}}}
\caption{Plots of both sides of Eq.~(\ref{F-15}) (white Cauchy
noise): function $y_1$ (solid curves), function $y_2$ (dashed
curve), for three values of time, namely: $t = 1,\, 1.5,\, 2.493$.
The critical time is $t_c = 2.493$ (black blue curve). The values of
the other parameters are: $x_0=0.2$, $r=1$, $D_1=0.5$.}
\label{fig-6}
\end{figure}
\begin{figure}[ptbh] \vspace{5mm}
\centering{\resizebox{7cm}{!}{\includegraphics{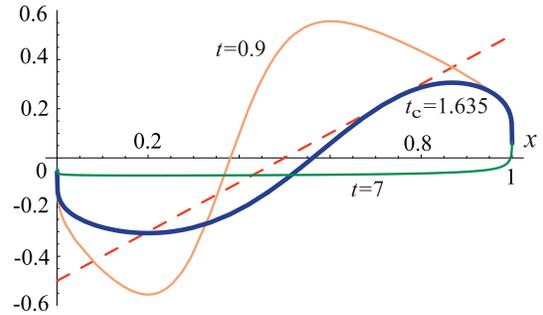}}}
\caption{Plots of both sides of Eq.~(\ref{F-15}) (white Cauchy
noise): function $y_1$ (solid curves), function $y_2$ (dashed
curve), for three values of time, namely: $t = 0.5,\, 1,\, 1.635$.
The critical time is $t_c = 1.635$ (black blue curve). The values of
the other parameters are: $x_0=0.2$, $r=1$, $D_1=1$.} \label{fig-7}
\end{figure}
The three values of the critical time $t_c$ corresponding to the
three values of the L\'{e}vy noise intensity investigated are: $D_1
= 0.5,~t_c = 2.493;~D_1 = 1,~t_c = 1.635;~D_1 = 5,~t_c = 0.247$. One
rough evaluation of the critical time $t_c$ is obtained by putting
equal to $1$ the scale parameter of the Cauchy distribution of
Eq.~(\ref{F-13}), that is $t_c \sim 1/D_{1}$. The critical time
$t_c$ is the time at which the maximum and one minimum of the
probability distribution (see Figs.~\ref{fig-3},~\ref{fig-4},
and~\ref{fig-5}) coalesce in one inflection point and in this point
$x$ the function $y_2 = x - 1/2$ becomes tangent at the function
$y_1$ (see Figs.~\ref{fig-6},~\ref{fig-7}, and~\ref{fig-8}). It is
interesting to note that the critical time $t_c$ decreases with the
noise intensity $D_1$. This is because by increasing the noise
intensity, more quickly the population density reaches the two
points near the boundaries $x = 0$ and $x = 1$.
\begin{figure}[ptbh] \vspace{5mm}
\centering{\resizebox{7cm}{!}{\includegraphics{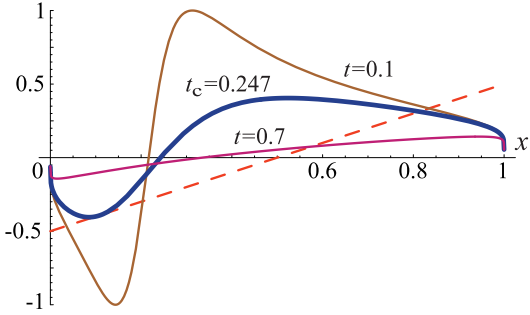}}}
\caption{Plots of both sides of Eq.~(\ref{F-15}) (white Cauchy
noise): function $y_1$ (solid curves), function $y_2$ (dashed
curve), for three values of time, namely: $t = 0.5,\, 1,\, 1.635$.
The critical time is $t_c = 0.247$ (black blue curve). The values of
the other parameters are: $x_0=0.2$, $r=1$, $D_1=5$.} \label{fig-8}
\end{figure}

\section{Nonlinear relaxation time of the mean population density}

It must be emphasized that to find the time evolution of the mean
population density one can use two different approaches. The first
one was proposed in Ref.~\cite{Jac89}. According to the exact
solution~(\ref{F-7}) of the Verhulst equation~(\ref{F-4}), we can
rewrite this expression in the following form
\begin{equation}
x\left(t\right) = f\left(e^{-rt-L\left(t\right)}\right),
\label{F-17}
\end{equation}
where
\begin{equation}
f\left(q\right)=\left(1+\frac{1-x_{0}}{x_{0}}q(t) \right)^{-1}.
\label{F-18}
\end{equation}
Then, by expanding the smooth function~(\ref{F-18}) in a standard
Taylor power series in $q$ around the point $q=0$ we have
\begin{equation}
f\left(q\right) = \sum\limits_{n=0}^{\infty}\frac{f^{\left(
n\right)} \left(0\right)}{n!}q^{n}.
\label{F-19}
\end{equation}
After substitution of Eq.~(\ref{F-19}) in Eq.~(\ref{F-17}) and
averaging we obtain
\begin{equation}
\left\langle x\left(t\right) \right\rangle =
\sum\limits_{n=0}^{\infty} \frac{f^{\left(n\right)} \left(
0\right)e^{-nrt}}{n!}\left\langle e^{-nL\left(t\right)}\right\rangle
\label{F-20}
\end{equation}
or, in accordance with Eq.~(\ref{F-8}),
\begin{eqnarray}
\left\langle x\left(  t\right)  \right\rangle
&=&\sum\limits_{n=0}^{\infty} \frac{f^{\left(  n\right)  }\left(
0\right) e^{-nrt}}{n!}\label{F-21} \\
&\times&\exp\left\{ t\int_{-\infty}^{+\infty}\frac{e^{-nz}-1+n\sin
z}{z^{2}}\,\rho\left( z\right) dz\right\} .\nonumber
\end{eqnarray}
For white Gaussian noise $\xi\left(t\right)$ with kernel
$\rho\left(z\right) = 2D \delta(z)$ we obtain from Eq.~(\ref{F-21})
the following asymptotic series
\begin{equation}
\left\langle x\left(t\right) \right\rangle
=\sum\limits_{n=0}^{\infty}
\frac{f^{\left(n\right)}\left(0\right)}{n!}e^{Dtn^{2}-nrt}.
\label{F-22}
\end{equation}
By considering a finite number of terms in this expansion leads to a
wrong conclusion about the critical slowing down phenomenon in such
a system, as found in Ref.~\cite{Leu88}. The exact result is
obtained, of course, by summing all the terms in Eq.~(\ref{F-22}).
Moreover, for most of the kernels $\rho(z)$ the integral in
Eq.~(\ref{F-21}) diverges. Thus, this approach is inappropriate for
our purposes, and it is better to use the direct average in
Eq.~(\ref{F-7}). Therefore, using this second approach we have
\begin{equation}
\left\langle x\left(  t\right)  \right\rangle
=\int\nolimits_{-\infty }^{+\infty}\left(
1+\frac{1-x_{0}}{x_{0}}e^{-rt-z}\right)  ^{-1}P_{L}\left(
z,t\right)dz.
\label{F-23}
\end{equation}

\begin{figure}[ptbh] \vspace{5mm}
\centering{\resizebox{7cm}{!}{\includegraphics{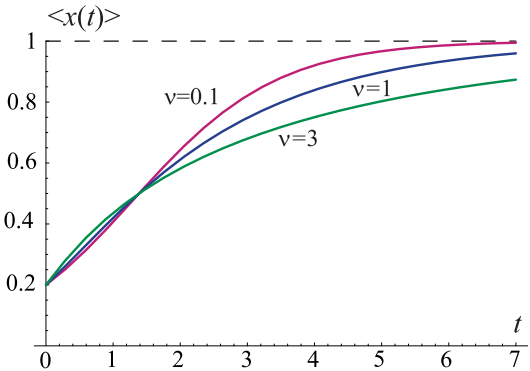}}}
\caption{Nonlinear relaxation of the mean population density in the
case of white shot noise excitation, for three values of the mean
frequency $\nu$, namely $\nu = 0.1,1,3$. The values of the other
parameters are: $x_0 = 0.2, r = 1, a_0 = 1$.} \label{fig-9}
\end{figure}

Let us consider now different models of white non-Gaussian noise
$\xi\left(t\right)$. We start with the white shot noise
\begin{equation}
\xi\left(t\right) = \sum\limits_{i}a_{i}\delta\left(t-t_{i}\right)
\label{F-24}
\end{equation}
having the symmetric dichotomous distribution of the pulse amplitude
$P(a) = \left[\delta\left(a-a_{0}\right) + \delta
\left(a+a_{0}\right)\right] /2$, mean frequency $\nu$ of pulse
train, and kernel $\rho\left(z\right) = \nu z^{2} P\left(z\right)$.
From Eq.~(\ref{F-8}) we have
\begin{equation}
\left\langle e^{iuL\left(t\right)}\right\rangle = e^{-\nu
t\left(1-\cos a_{0}u\right)}.
\label{F-25}
\end{equation}
By making the reverse Fourier transform in Eq.~(\ref{F-25}) we find
the probability distribution of the corresponding L\'{e}vy process
\begin{equation}
P_{L}\left( z,t\right) =e^{-\nu
t}\sum_{n=-\infty}^{+\infty}I_n\left( \nu t\right) \delta \left(
z-na_0\right) ,
\label{F-26}
\end{equation}
where $I_n\left( x\right) $ is the $n$-order modified Bessel
function of the first kind. The relaxation of the mean population
density $x(t)$ is shown in Fig.~\ref{fig-9}. According to the
Eq.~(\ref{F-23}) and~(\ref{F-26}) the stationary value of the
population density in such a case is $\left\langle x\right\rangle
_{st}=1$, but the relaxation time~(\ref{F-3}) increases with
increasing the mean frequency of pulses.

\begin{figure}[ptbh] \vspace{5mm}
\centering{\resizebox{7cm}{!}{\includegraphics{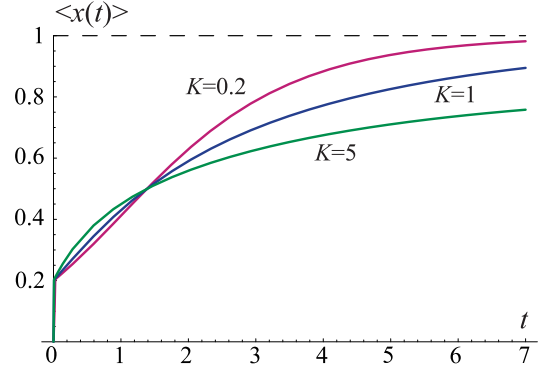}}}
\caption{Nonlinear relaxation of the mean population density in the
case of L\'{e}vy noise with distribution~(\ref{F-13}), for three
values of the parameter $K$, namely $K = 0.2,1,5$. The values of the
other parameters are: $x_0 = 0.2, r = 1$.} \label{fig-10}
\end{figure}

For white non-Gaussian noise with the kernel $\rho(z) =
Kz/\left(2\sinh z\right)$ we observe a similar transient dynamics,
which is shown in Fig.~\ref{fig-10}. We have the same stationary
value $\left\langle x\right\rangle_{st}$, and the relaxation time
$T$ increases with increasing the parameter $K$, which is
proportional to the noise intensity.

\begin{figure}[ptbh] \vspace{5mm}
\centering{\resizebox{7cm}{!}{\includegraphics{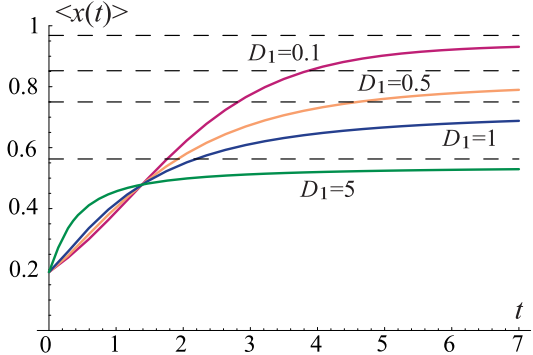}}}
\caption{Nonlinear relaxation of the mean population density in the
case of white Cauchy noise, for four values of the noise intensity
$D_1$, namely $D_1 = 0.1,0.5,1,5$. The values of the other
parameters are: $x_0 = 0.2, r = 1$.} \label{fig-11}
\end{figure}

Finally, in the case of white Cauchy noise $\xi\left(t\right)$ we
obtain interesting exact analytical results. First of all,
substituting Eq.~(\ref{F-13}) in Eq.~(\ref{F-23}) and changing the
variable $z=D_{1}ty$ under the integral, we obtain
\begin{equation}
\left\langle x\left(t\right) \right\rangle = \frac{1}{\pi}\int
\nolimits_{-\infty}^{+\infty}\left[
1+\frac{1-x_{0}}{x_{0}}e^{-t\left( r+D_{1}y\right)}\right]
^{-1}\frac{dy}{1+y^{2}}.
\label{F-27}
\end{equation}
For the stationary mean value $\left\langle x\right\rangle _{st}$ we
find from Eq.~(\ref{F-27})
\begin{equation}
\left\langle x\right\rangle
_{st}=\lim_{t\rightarrow\infty}\left\langle x\left(  t\right)
\right\rangle =\frac{1}{\pi}\int\nolimits_{-\infty
}^{+\infty}\frac{1\left(  r+D_{1}y\right)  dy}{1+y^{2}},
\label{F-28}
\end{equation}
where $1\left(x\right)$ is the step function. After evaluation of
the integral in Eq.~(\ref{F-28}) we obtain finally
\begin{equation}
\left\langle x\right\rangle
_{st}=\frac{1}{2}+\frac{1}{\pi}\arctan\frac {r}{D_{1}}.
\label{F-29}
\end{equation}
As it is seen from Fig.~\ref{fig-11} and Eq.~(\ref{F-29}), for small
noise intensity $D_{1}$, with respect to the value of the rate
parameter $r = 1$, the stationary mean value of the population
density is approximately $1$, as for the other white non-Gaussian
noise excitations considered. But for large values of $D_{1}$, this
asymptotic value, which is independent from the initial value of
population density $x_{0}$, tends to $0.5$.

\begin{figure}[ptbh] \vspace{5mm}
\centering{\resizebox{7cm}{!}{\includegraphics{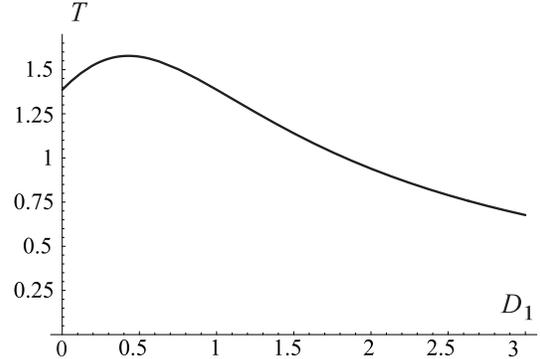}}}
\caption{Nonmonotonic behavior of the nonlinear relaxation time $T$
as a function of the white Cauchy noise intensity $D_1$. The values
of the other parameters are: $x_0 = 0.5, r = 1$.} \label{fig-12}
\end{figure}

It is interesting also to analyze, for this case of white Cauchy
noise, the dependence of the relaxation time $T$ from the noise
intensity $D_{1}$. Substituting Eq.~(\ref{F-27}) in Eq.~(\ref{F-3})
and changing the order of integration, for initial condition $x_0 =
0.5$, we are able to calculate analytically the double integral in
$t$ and in $y$ obtaining the final result
\begin{equation}
T=\frac{\pi\ln2}{r\left(  1+D_{1}^{2}/r^{2}\right) arccot(D_{1}/r)}.
\label{F-30}
\end{equation}
We find a non-monotonic behavior of the relaxation time $T$ versus
the noise intensity $D_{1}$ with a maximum at the noise intensity
$D_1 = 0.43$, as shown in Fig.~\ref{fig-12}.
This nonmonotonic behavior is also visible for another initial
position $x_0 = 0.2$, as shown in Fig.~\ref{fig-11}. Here the
relaxation time to reach the stationary value of population density
$x_{st}$ increases from very low noise intensity ($D_1 = 0.1$) to
moderate low intensity ($D_1 = 0.5$), while decreases for higher
noise intensities ($D_1 = 1.5$). This is also due to the dependence
of $x_{st}$ from the noise intensity $D_1$ (see Eq.~(\ref{F-30})).
We note that this non-monotonic behavior of the relaxation time $T$
is related to the peculiarities of the transient dynamics of the
mean population density and it will be object of further
investigations.

\section{Conclusions}

The transient dynamics of the Verhulst model, perturbed by arbitrary
non-Gaussian white noise, is investigated. This well-known equation
is an appropriate ecological and biological model to describe
closed-population dynamics, self-replication of ~macromolecules
under constraint, cancer growth, spread of viral epidemics, etc...
By using the properties of the infinitely divisible distribution of
the generalized Wiener process, we analyzed the effect of different
non-Gaussian white sources on the nonlinear relaxation of the mean
population density and on the time evolution of the probability
distribution of the population density. We obtain exact results for
the nonstationary probability distribution in all cases investigated
and for the Cauchy stable noise we derive the exact analytical
expression of the nonlinear relaxation time. Due to the presence of
a L\'{e}vy multiplicative noise, the probability distribution of the
population density exhibits a transition from a trimodal to a
bimodal distribution in asymptotics. This transition, characterized
by the appearance of a minimum, happens at a critical time $t_c$,
which can be roughly evaluated as $t_c \sim 1/D_1$ (where $D_1$ is
the noise intensity) and exactly evaluated from the
condition~(\ref{F-15}). Finally a nonmonotonic behavior of the
nonlinear relaxation time of the population density as a function of
the Cauchy noise intensity was found.

\vspace{-0.6cm}
\section*{Acknowledgements}

We acknowledge support by MIUR, CNISM-INFM, and Russian Foundation
for Basic Research (project 08-02-01259).

\end{document}